\documentclass[10pt,conference]{IEEEtran}

  \usepackage{graphicx}
  \graphicspath{{eps/}}
  \DeclareGraphicsExtensions{.eps}

\usepackage{subfigure} 

\usepackage{url}        

\usepackage{amsmath}    
\usepackage[all]{xy}
\usepackage{amssymb}
\usepackage[footnotesize]{caption}
\usepackage{comment}

\newenvironment{mylist}[1][$\bullet$]{\begin{list}{#1}{\itemindent .5\parindent \leftmargin 0mm }}{\end{list}}
\newcommand{\myitem}{\item}

\newtheorem{theo}{Theorem}
\newtheorem{prop}[theo]{Proposition}

\newtheorem{defi}[theo]{Definition}
\newtheorem{rema}[theo]{Remark}

\renewcommand{\H}{{\cal H}}
\newcommand{\T}{{\cal T}}
\newcommand{\N}{{\cal N}}
\newcommand{\M}{{\cal M}}
\newcommand{\C}{{\cal C}}

\renewcommand{\P}{{\cal P}}
\newcommand{\Q}{{\cal Q}}
\newcommand{\X}{{\cal X}}
\newcommand{\fmin}{{f_{\mbox{\scriptsize min}}}}
\newcommand{\fmax}{{f_{\mbox{\scriptsize max}}}}
\newcommand{\dcmax}{d^{\mbox{\scriptsize c}}\!\!{\raisebox{0mm}{\rm\tiny max}}}
\newcommand{\leaf}{\mbox{\rm Leaf}}

\begin{document}

\title{Iterative LDPC decoding using neighborhood reliabilities}

\author{
\authorblockN{Valentin Savin}
\authorblockA{CEA-LETI, MINATEC, 17 rue des Martyrs, 38054 Grenoble, France \\
valentin.savin@cea.fr}
}

\maketitle

\begin{abstract}
In this paper we study the impact of the processing order of nodes of a bipartite graph, on the performance
of an iterative message-passing decoding. To this end, we introduce the concept of {\em neighborhood
reliabilities} of graph's nodes. Nodes reliabilities are calculated at each iteration and then are used to
obtain a processing order within a serial or serial/parallel scheduling. The basic idea is that by processing
first the most reliable data, the decoder is reinforced before processing the less reliable one. Using
neighborhood reliabilities, the Min-Sum decoder of LDPC codes approaches the performance of the Sum-Product
decoder.
\end{abstract}


%
\IEEEpeerreviewmaketitle

\section{Introduction}
\label{sec:scheduled_iter_decoding}

Low density parity check codes can be iteratively decoded using a parallel, serial or mixed (serial/parallel)
scheduling. The parallel scheduling consists in processing all check nodes of the associated Tanner graph and
then update all variable-to-check messages. The serial scheduling may apply to variable or check nodes.
Applied to check nodes, it consists in progressively processing each check node and updating outgoing
variable-to-check messages from all variable nodes connected to the processed check node. Equivalently, each
check node processing is preceded by the update of the incoming variable-to-check messages. The mixed
scheduling can be described as a serial scheduling applied to groups of check or variable nodes, which
performs a parallel scheduling inside each group of nodes. Several serial or mixed schedules where proposed
in the
literature \cite{bout-tous-guil}, \cite{Mao-Bani}, \cite{Shar-Lits-Gold}, \cite{zhan-foss}. 
The only advantage of the serial scheduling reported in the literature is that the decoding algorithm
converges about twice faster as when the parallel scheduling is used \cite{Shar-Lits-Gold}, \cite{zhan-foss}.
Due to this fact, the serial scheduling also performs better than the parallel scheduling when a very small
number of iterations, {\em e.g.} $10$ iterations, is used \cite{bout-tous-guil}, \cite{zhan-foss}.

In this paper, we focus on the impact of different scheduling schemes on the decoder performance, assuming a
large number of iterations. By concern of brevity, the paper addresses only the case of binary LDPC codes,
but the presented results also apply to non-binary codes. We fix the following notations:

\begin{mylist}
\myitem $\H$ is a Tanner graph comprising $N$ variable nodes and $M$ check nodes, 
\myitem $\N = \{n \mid 1\leq n \leq N\}$ is the set of variable nodes, 
\myitem $\M = \{m \mid 1 \leq n \leq M\}$ is the set of check nodes, 
\item $\gamma_n$ and $\tilde{\gamma}_n$ are the a priori, respectively a posteriori,
information available at the variable node $n$,
\item $\alpha_{m,n}$ is the variable-to-check message from $n$ to  $m$,
\item $\beta_{m,n}$, is the check-to-variable message from $m$ to $n$.
\end{mylist}

The analysis of iterative decoders is generally a challenging task. This is mainly due to cycles appearing in
the Tanner graph, which depreciate the extrinsic nature of the exchanged information. One approach that is
less sensitive to graph's cycles is the use of {\em computation trees} associated with the decoding process
\cite{Wiberg}. Fix some variable node $n$ and consider the a posteriori information $\tilde \gamma_n$. By
examining the updates that have occurred, one may recursively trace back through time the computation of
$\tilde \gamma_n$. This trace back will form a tree graph rooted at $n$ and consisting of interconnected
variable and check nodes in the same way as in the original graph, but the same variable or check nodes may
appear at several places in the tree. The fundamental observation is that the computation tree does not
depend only on the Tanner graph or the scheduling type but also on the updating order. The success of
decoding the variable node $n$ depends on the reliability of the information conveyed through the computation
tree.

Let us be more specific. Thereafter we assume that the all-zero codeword is transmitted through the
channel 
 (for symmetric channels this assumption does not induce any loss of generality). We also assume that
 the $0$ bit corresponds to a positive information. Let $\T$ denote the
computation tree of $\tilde\gamma_n$ -- the variable node $n$ will remain fixed through this section. We also
note by $\P$ and $\Q$ the sets of variable and respectively check nodes of $\T$. 
The set $\P$ has a distinguished
element, namely the tree's root, which will be still denoted by $n$.
A $\T$-codeword is a vector $(x_{p})_{p\in\P}$ satisfying all constraints corresponding to check nodes
$q\in\Q$. Its {\em multiplicity vector} is an integer-valued vector $([x]_k)_{k\in\N}$, where $[x]_k$ is the
number of variable nodes $p\in\P$, corresponding to the variable node $k\in\N$, such that $x_p = 1$.

We recall (see \cite{Wiberg}) that a $\T$-codeword $(x_{p})_{p\in\P}$ such that $x_n = 1$ is called a {\em
minimal deviation} if it does not cover any other nonzero $\T$-codeword.
In \cite{Wiberg}, theorem $4.2$, it is proved that when the {\bf Min-Sum} decoding is used, a necessary
condition for a decoding error (on bit $n$) to occur is that for some minimal deviation $x$ we have:

\begin{equation}\label{eq:1}
\sum_{k\in\N}[x]_k\gamma_k \leq 0
\end{equation}

Obviously, the probability that such an inequality holds increases when the computation tree contains several
copies $p$ of some {\em unreliable} variable nodes $k$; that is, when there exist minimal deviations $x$
having large $[x]_k$ for variable nodes $k$ with $\gamma_k \leq 0$.

For a serial or mixed scheduling, check nodes that are processed first and consequently, variable nodes connected to
them, have a large number of copies appearing in the computation tree. From the above discussion, we conclude
that the decoder should process first the most reliable check nodes. For this, we have to specify what a {\em
reliable} check node should be. Intuitively, the reliability of a check node should be a measure of the
reliabilities of variable nodes connected to it or, more involved, a measure of the reliabilities of variable
nodes located in some neighborhood of a specified depth. This will be formalized in the following section
where we introduce neighborhood reliabilities. However, it should be clear that anyhow we will define the
reliability of a check node, its estimation should be updated at each iteration, rather than be computed only
once before starting decoding.

\section{Neighborhood reliabilities}
\label{sec:reliability}

The {\em depth-$d$ neighborhood} of a node $\pi$ of $\H$, denoted by ${\cal V}_\pi^{(d)}$, is the set
 of nodes whose distance to $\pi$ is lower or equal to $d$.

When a message passing iterative decoding exchanges messages along the edges of the Tanner graph, we consider
that the following {\em soft} information is available at graph's nodes:
\begin{list}{$\bullet$}{\itemindent \parindent \leftmargin 0mm}
\item the information available at variable nodes, which consists of the a priori information, the
a posteriori information and the variable-to-check messages,
\item the information available at check nodes consisting of check-to-variable messages.
\end{list}
Moreover, at each iteration and for each variable node $n$, a hard bit $s_n$ is computed according to the a
posteriori information $\tilde{\gamma}_n$. For each check node $m$ we also consider the parity check value
$\sigma_m = \displaystyle \bigoplus_{n\in{\cal H}(m)} s_n \in \{0,1\}$ (sum $\mod 2$). This value constitutes
the {\em hard} information available at the check node $m$. The check node $m$ is said to be {\em verified}
if $\sigma_m = 0$.
\begin{defi}
We call {\em neighborhood reliability of depth $d$}, or simply {\em reliability of depth $d$}, a function
associating to any node $\pi$ a real value depending on the information available at nodes $\tau \in {\cal
V}_\pi^{(d)}$ and that is used to determine a processing order during a message passing iterative decoding.
\end{defi}

When nodes $\pi$ from the above definition are only either of check or variable type, we speak about
reliability functions for check, respectively variable, nodes.

We notice that the amplitude of a soft information represents a measure of its reliability. The hard information
$\sigma_m$ can be used to detect whenever non accurate information is conveyed through the graph. Let us
detail this point.
Let $m$ be an unverified check node; meaning that $\sigma_m = 1$ after a certain number, say $l$, of iterations of the
Min-Sum decoding. Using the all-zero codeword assumption, it follows that there is a variable node $n\in
\H(m)$ such that $\tilde\gamma_n < 0$ (we assume that the $0$ bit corresponds to a positive information). From
the theorem $4.2$ in \cite{Wiberg}, it follows that, at iteration $l$, the computation tree of $\tilde\gamma_n$ contains a
minimal deviation satisfying (\ref{eq:1}). If, at the next iteration, unverified check nodes are processed first, many copies of such
minimal deviations will appear as subtrees of different a posteriori information computation trees. Therefore, such a processing order would
increase the probability of minimal deviations satisfying (\ref{eq:1}) to appear in computation trees at iteration $l+1$. Here, we also use the fact that
any minimal deviation of a subtree extends to a minimal deviation of the global tree (this can be
easily derived using the recursive construction of minimal deviations described in section
\ref{sec:cutting_back}).


\subsection{Examples}
\label{subsec:reliability:exemples}

We restrict ourself only to reliability functions for check nodes.
 In order to obtain better reliability functions, we have to consider deeper neighborhoods. Obviously, increasing the
 neighborhoods depth, results in an increased complexity of reliabilities computation and can create problems with handling the
 cycles of the Tanner graph. In practice, depth-$2$ reliability functions seem to be a good compromise. We
 define a real-valued, depth-$2$, check nodes reliability function, $f(m)$, as follows. First, we define
 $f'(m)$ to be the sum of reliabilities of the soft information available at the unverified check nodes
 $m'\in{\cal V}^{(2)}_m\setminus \{m\}$; so, assuming the graph is $4$-cycle free, we have:
 $$f'(m) = \sum_{n\in\H(m)}\left(\sum_{\stackrel{m'\in\H(n)\setminus\{m\}}{\sigma_{m'} = 1}}\mid\beta_{m',n}\mid\right)$$
As discussed above, the only circumstance when we can detect an unreliable information conveyed through a
check node is when the check node is unverified. That explains why we consider only unverified check nodes in
the above definition. Thus, the function $f'(m)$ gives an extrinsic measure of the unreliability, rather than
the reliability, of a check node $m$. This means that a check node $m_1$ is more reliable than a check node
$m_2$ if $f'(m_1) < f'(m_2)$. To prevent that an unverified check node $m_1$ be more reliable than a verified
check node $m_2$, we define the function $f(m)$ taking into account both $\sigma_m$ and $f'(m)$, as follows:
$$f(m) = (\sigma_m, f'(m))$$
Consequently, we say that a check node $m_1$ is more reliable than a check node $m_2$ if $\sigma_{m_1} <
\sigma_{m_2}$ (thus, $m_1$ is verified while $m_2$ is not) or if $\sigma_{m_1} = \sigma_{m_2}$ and $f'(m_1) <
f'(m_2)$. This corresponds to the lexicographical order between $f(m_1)$ and $f(m_2)$. In this case, check
nodes can be processed using a serial (or mixed) scheduling, starting with the most reliable check node(s)
and ending with the less reliable one(s).

As we will explain shortly, integer-valued functions considerable reduce the use cost of neighborhood
reliabilities. 
Therefore, we propose an integer-valued reliability function $\bar f(m)$, as an alternative to the
real-valued function defined above. First, for each check node $m$ and variable node $n\in\H(m)$ we define
$\sigma_{m,n}$ as a {\em flag} indicating whenever there are or not unverified check nodes
$m'\in\H(n)\setminus\{m\}$. That is $\sigma_{m,n} = 0$ if $\sigma_{m'} = 0$  for all $
m'\in\H(n)\setminus\{m\}$ and $\sigma_{m,n} = 1$ otherwise.
%
Then, we define $f(m)$ as follows:
$$\bar f(m) = \sigma_m(\dcmax+1) + \displaystyle \sum_{n\in {\cal H}(m)} \sigma_{m,n} $$
where $\dcmax$ is the maximum check degree. We note that the sum $\displaystyle \sum_{n\in {\cal H}(m)}
\sigma_{m,n}$ increments the value of $\bar f(m)$ by $1$ for each variable node $n\in {\cal H}(m)$ that is
connected to at least one unverified check node, other than $m$. The minimum of this sum is $0$ and it is
obtained if all check nodes (except $m$) located in the depth-$2$ neighborhood of $m$ are verified. Its
maximum is equal to $\dcmax$ and it is obtained if all variable nodes $n\in \H(m)$ are connected to at least
one unverified check node, other than $m$. Thus, we declare a check node $m_1$ more reliable than a check
node $m_2$ if $\bar f(m_1) \leq \bar f(m_2)$. The role of the term $\sigma_m(\dcmax+1)$ is to make sure that
a verified check node will always be declared more reliable than an unverified one. Precisely, we have $\bar
f(m) \in \{ 0, \dots, \dcmax\}$ if $m$ is verified and $\bar f(m) \in \{ \dcmax + 1, \dots, 2\dcmax + 1 \}$
if it is not.

\subsection{Use of neighborhood reliabilities}
\label{subsec:reliability:use}

As in the above section, we consider only reliability functions for check nodes. The description below also
applies to reliability functions for variable nodes, by reversing the roles of variable and check nodes. We
distinguish two cases, according to whether the reliability function is real or integer-valued.

\subsubsection{Real-valued reliability functions}\label{subsubsec:reliability:use:real}
\begin{figure}[!t]
\centering
\includegraphics[scale=0.85,clip=true]{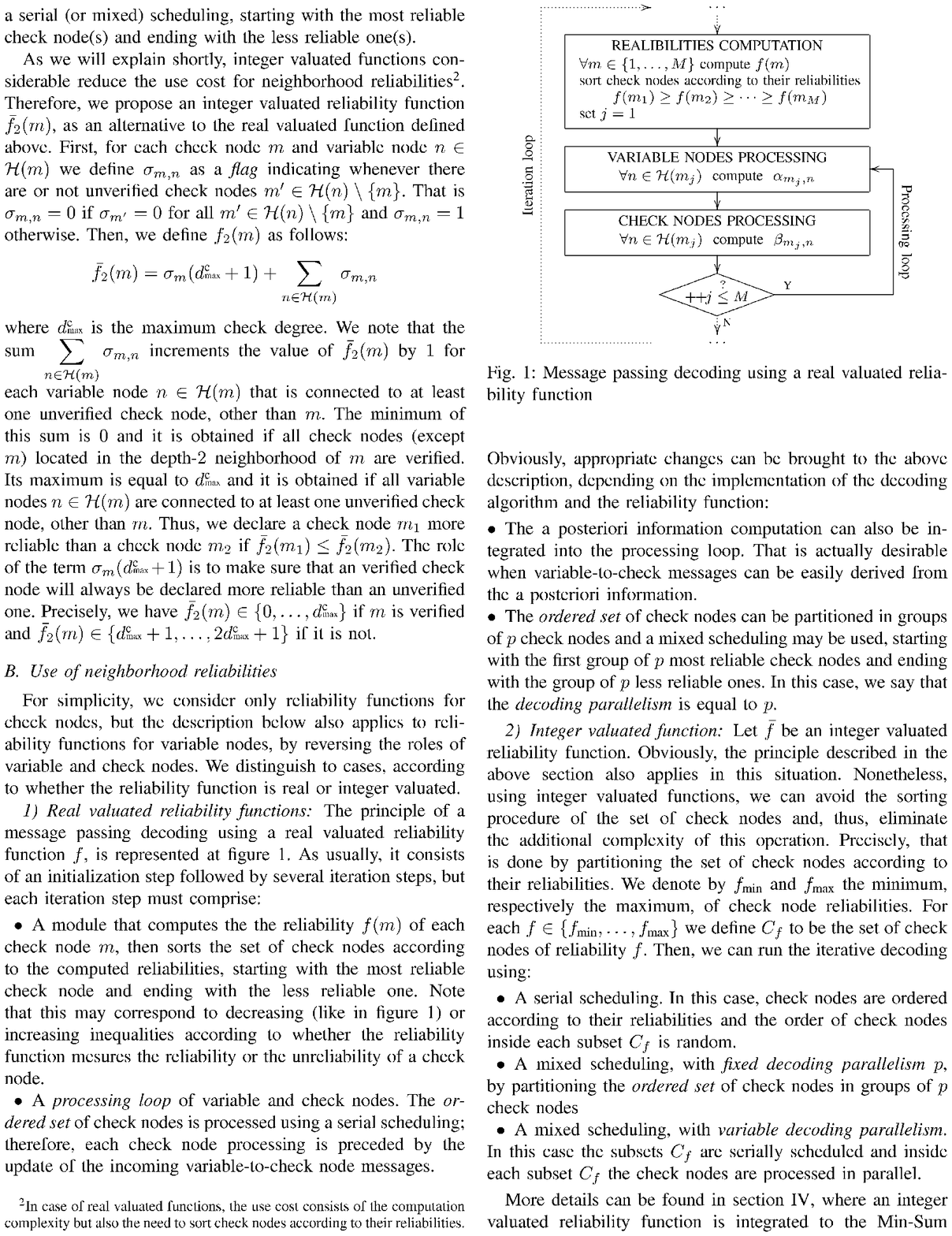}
\vspace{-2mm} \caption{Iterative decoding using a real-valued reliability function}\vspace{-5mm}
\label{fig:decoding_with_real_reliability}
\end{figure}

 The principle of a message passing decoding using a real-valued reliability function $f$,
 is represented at figure \ref{fig:decoding_with_real_reliability}.
 As usually, it consists in an initialization step followed by several iteration steps, each one of which includes:

\begin{list}{$\bullet$}{\itemindent \parindent \leftmargin 0mm}
\item A module that computes the reliability $f(m)$ of each check node $m$. The set of check nodes is sorted
according to the computed reliabilities, starting with the most reliable check node and ending with the less
reliable one. This may correspond to decreasing (as in figure \ref{fig:decoding_with_real_reliability}) or
increasing inequalities according to whether the reliability function mesures the reliability or the
unreliability of a check node.
\item A {\em processing loop} of variable and check nodes. The {\em ordered set} of check nodes is processed using a
serial scheduling. Therefore, each check node processing is preceded by the update of the incoming
variable-to-check messages.
\end{list}
Obviously, appropriate changes can be brought to the above description, depending on the implementation of
the decoding algorithm and the reliability function:
 \begin{mylist}
 \item The computation of the a posteriori information can also be integrated into the
 processing loop. This is actually desirable when variable-to-check messages can be easily derived
 from the a posteriori information.
 \item The {\em ordered set} of check nodes can be partitioned in groups of $p$ check nodes and a mixed scheduling may be used,
 starting with the first group of $p$ most reliable check nodes and ending with the group of $p$ less
 reliable ones. In this case, we say that the {\em decoding parallelism} is equal to $p$.
\end{mylist}

\subsubsection{Integer-valued function}\label{subsubsec:reliability:use:int}
Let $\bar f$ be an integer-valued reliability function. Obviously, the principle described in the above
section also applies in this situation. Nonetheless, using integer-valued functions, one can avoid sorting
the set of check nodes and, thus, eliminate the additional complexity of this operation. Precisely, that is
done by partitioning the set of check nodes according to their reliabilities. We denote by $\fmin$ and
$\fmax$ the minimum, respectively the maximum, of check node reliabilities. For each
$f\in\{\fmin,\dots,\fmax\}$ we define $C_f$ to be the set of check nodes of reliability $f$. Then, we can run
the iterative decoding using:

\begin{list}{$\bullet$}{\itemindent \parindent \leftmargin 0mm}
\item A serial scheduling. In this case, check nodes are ordered according to their reliabilities and
the order of check nodes inside each subset $C_f$ is random.
\item A mixed scheduling, with {\em fixed decoding parallelism} $p$, by partitioning the {\em ordered set} of
check nodes in groups of $p$ check nodes
\item A mixed scheduling, with {\em variable decoding parallelism}.
In this case the subsets $C_f$ are serially scheduled and inside each subset $C_f$ the check nodes are
processed in parallel.
\end{list}

More details can be found in section \ref{sec:min_sum_with_reliability}, where an integer-valued reliability
function is integrated to the Min-Sum decoding of LDPC codes.

\section{Cutting back the computation tree}
 \label{sec:cutting_back}
In section \ref{sec:scheduled_iter_decoding} we explained how the processing order of check nodes affects
minimal deviations of computation trees and therefore impacts the decoding performance. That motivated the
introduction of neighborhood reliabilities in section \ref{sec:reliability}. In this section we show that it
is possible to {\em force minimal deviations to pass through reliable nodes} of the computation tree.

\begin{figure}[!t]
\centering
\includegraphics[scale=0.85,clip=true]{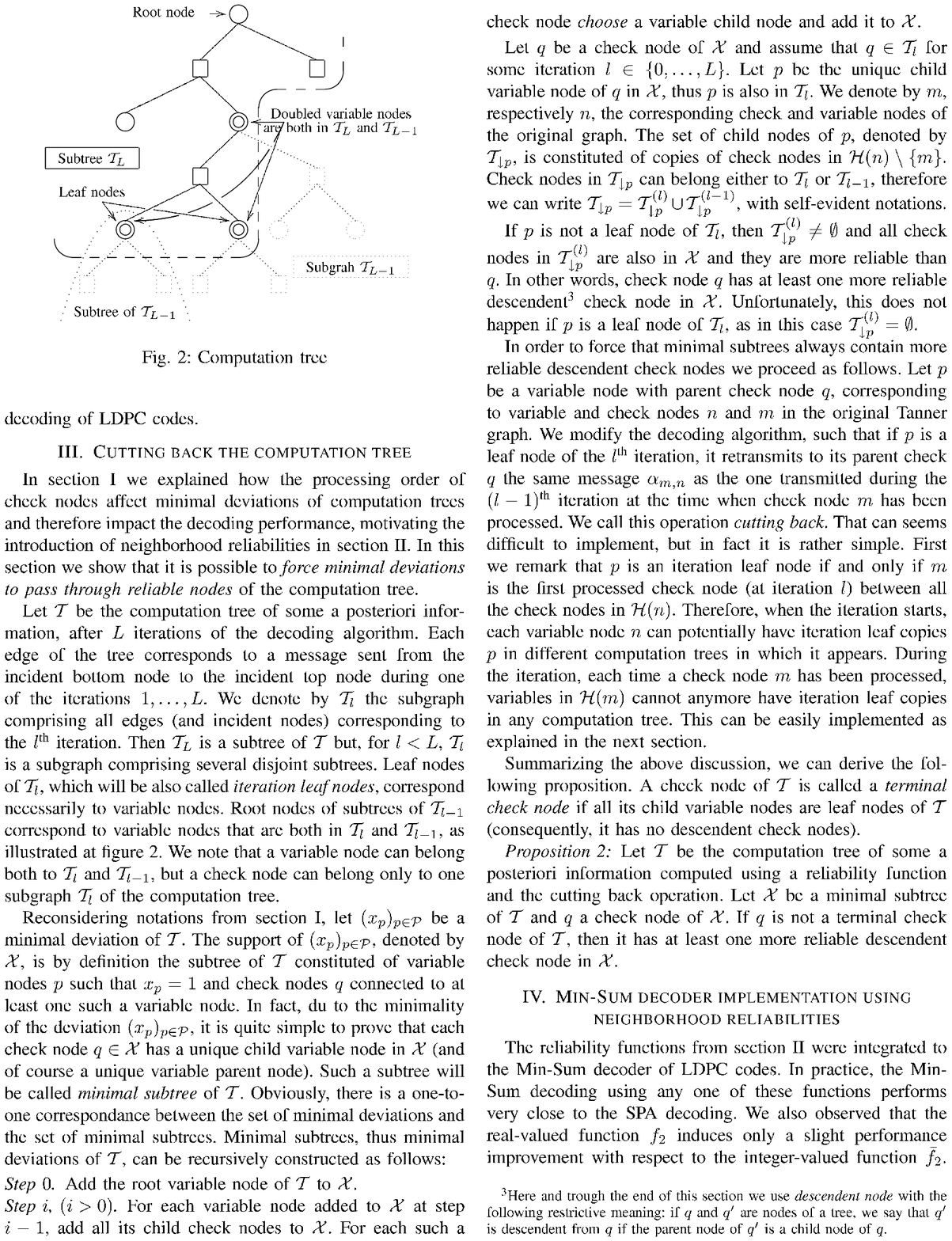}
\vspace{-2mm}\caption{Computation tree} \label{fig:computation_tree}\vspace{-5mm}
\end{figure}

Let $\T$ be the computation tree of some a posteriori information, after $L$ iterations of the decoding
algorithm. Each edge of the tree corresponds to a message sent from the incident bottom node to the incident
top node during one of the iterations $1,\dots,L$. We denote by $\T_l$ the subgraph comprising all edges (and
incident nodes) corresponding to the $l^{\mbox{\scriptsize th}}$ iteration. Then $\T_L$ is a subtree of $\T$
while, for $l < L$, $\T_{l}$ is a subgraph comprising several disjoint subtrees. Leaf nodes of $\T_l$, which
will be also called {\em iteration leaf nodes}, correspond necessarily to variable nodes.  Root nodes of
subtrees of $\T_{l-1}$ correspond to variable nodes that are both in $\T_l$ and $\T_{l-1}$, as illustrated in
figure \ref{fig:computation_tree}.

Reconsidering notations from section \ref{sec:scheduled_iter_decoding}, let $(x_p)_{p\in\P}$ be a minimal
deviation of $\T$. The support of $(x_p)_{p\in\P}$, denoted by $\X$, is by definition the subtree of $\T$
constituted of variable nodes $p$ such that $x_p = 1$ and check nodes $q$ connected to at least one such a
variable node. In fact, du to the minimality of the deviation $(x_p)_{p\in\P}$, it is quite simple to prove
that each check node $q\in\X$ has a unique child variable node in $\X$ (and of course a unique variable
parent node). Such a subtree will be called {\em minimal subtree} of $\T$. Obviously, there is a one-to-one
correspondance between the set of minimal deviations and the set of minimal subtrees. Minimal subtrees, thus
minimal deviations of $\T$, can be recursively constructed as follows:

\begin{mylist}
\item[{\em Step $0$.}] Add the root variable node of $\T$ to $\X$.
\item[{\em Step $i$, $(i>0)$.}] For each variable node added to $\X$ at step $i-1$, add all its child
check nodes to $\X$. For each such a check node, choose a variable child node and add it to $\X$.
\end{mylist}

Let $q$ be a check node of $\X$ and assume that $q\in\T_l$ for some iteration $l\in\{0,\dots,L\}$. Let $p$ be
the unique child variable node of $q$ in $\X$, thus $p$ is also in $\T_l$. We denote by $m$, respectively
$n$, the corresponding check and variable nodes of the original graph. The set of child nodes of $p$, denoted
by $\T_{\downarrow p}$, is constituted of copies of check nodes in $\H(n)\setminus\{m\}$. Check nodes in
$\T_{\downarrow p}$ can belong either to $\T_l$ or $\T_{l-1}$, therefore we can write $\T_{\downarrow p}=
\T^{(l)}_{\downarrow p} \cup \T^{(l-1)}_{\downarrow p}$, with self-evident notations.

If $p$ is not a leaf node of $\T_l$, then $\T^{(l)}_{\downarrow p} \not=\emptyset$ and all check nodes
$q'\in\T^{(l)}_{\downarrow p}$ are also in $\X$ and they are copies of check nodes $m'\in\H(n)$ that were processed at iteration $l$ before the check node $m$. Consequently, check node
$q$ has at least one more reliable descendent\footnote{Here and trough the end of this section we use {\em
descendent node} with the following restrictive meaning: if $q$ and $q'$ are nodes of a tree, we say that
$q'$ is descendent from $q$ if the parent node of $q'$ is a child node of $q$.} check node in $\X$.
Unfortunately, this does not happen if $p$ is a leaf node of $\T_l$, as in this case $\T^{(l)}_{\downarrow p}
= \emptyset$.

In order to force that minimal subtrees always contain more reliable descendent check nodes we 
modify the decoding algorithm as follows: if $p$ is a leaf node of the $l^{\mbox{\scriptsize th}}$ iteration,
it retransmits to its parent check $q$ the same message $\alpha_{m,n}$ as the one transmitted during the
$(l-1)^{\mbox{\scriptsize th}}$ iteration at the time when check node $m$ has been processed. We call this
operation {\em cutting back}.
Now, we remark that $p$ is an iteration leaf node if and only if $m$ is the first processed check node (at
iteration $l$) between all the check nodes in $\H(n)$. Therefore, when the iteration starts, each variable
node $n$ may potentially have iteration leaf copies $p$ in different computation trees in which it appears.
During the iteration, each time a check node $m$ has been processed, variable nodes in $\H(m)$ cannot anymore
have iteration leaf copies in any computation tree. This can be easily implemented as explained in the next
section.

Summarizing the above discussion, we can derive the following proposition. A check node of $\T$ is called a
{\em final check node} if all its child variable nodes are leaf nodes of $\T$ (consequently, it has no
descendent check nodes).

\begin{prop}
Let $\T$ be the computation tree of some a posteriori information computed using a reliability function and
the cutting back operation. Let $\X$ be a minimal subtree of $\T$ and $q$ a check node of $\X$. If $q$ is not
a final check node of $\T$, then it has at least one more reliable descendent check node in $\X$.
\end{prop}

\section{Min-Sum decoder implementation using neighborhood reliabilities}
\label{sec:min_sum_with_reliability}

The reliability functions from section \ref{subsec:reliability:exemples} were integrated to the Min-Sum
decoder of LDPC codes. In practice, the Min-Sum decoder using any one of these functions performs very close
to the SPA decoder. We also observed that the real-valued function $f$ induces only a slight performance
improvement with respect to the integer-valued function $\bar f$. Therefore, in this section we focus on the
implementation of the Min-Sum decoding using the integer-valued function $\bar f$, which has a lower
complexity. Nonetheless, we present simulation results for both integer and real-valued functions.

\begin{figure}[!t]
\centering
\includegraphics[scale=0.85,clip=true]{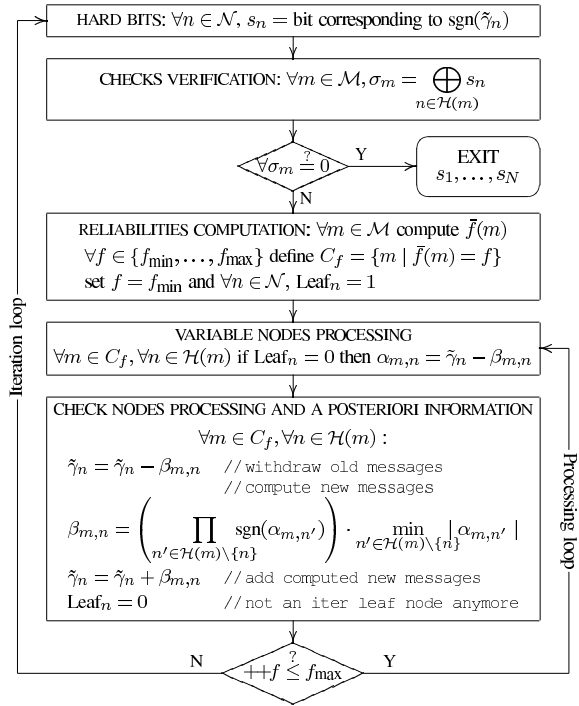}
\vspace{-2mm}\caption{Min-Sum decoding using an integer-valued reliability function}\vspace{-5mm}
\label{fig:int_reliability}
\end{figure}

\begin{mylist}
\item[{\em Initialization step}.] We initialize the a posteriori information $\tilde\gamma_n = \gamma_n$
and the check-to-variable messages $\beta_{m,n} = 0$.
\item[{\em Iteration loop}.] Its implementation is represented at figure
\ref{fig:int_reliability}. We use a mixed scheduling with variable decoding parallelism, as explained in
section \ref{sec:reliability}.

\begin{list}{$\bullet$}{\itemindent 3.5mm \leftmargin 0mm}
\item On top of the iteration loop we placed the hard bits computation and check nodes verification modules. This is
justified by the fact that the reliability function $\bar f$ makes use of parity check values $\sigma_m$.

\item They are followed by the reliabilities computation module. It computes check nodes reliabilities
and, accordingly, places check nodes in reliability sets $C_f$. Besides, it also set {\em leaf flags} of
variable nodes ($\leaf_n$) to $1$: at this moment any variable node can potentially have iteration leaf
copies in different computation trees in which it appears.

\item The {\em processing loop} is placed at the bottom of the of the iteration loop. Subsets $\C_f$ are
serially scheduled and check nodes inside one such a subset are processed in parallel. Each time that a new
check node $m$ is processed we update the a posteriori information of variable nodes $n\in\H(m)$. This is
done by withdrawing the old check-to-variable message, then computing the new one and finally adding the
new message to the a posteriori information, as it is shown in the {\em check nodes processing and a
posteriori information} module. Furthermore, each time a check node $m$ is processed we set leaf flags of
variable nodes $n\in\H(m)$ to $0$, as they cannot anymore have iteration leaf copies in any computation tree.
On top of this module we placed the {\em variable nodes processing} module. Thus, any variable-to-check message
can be updated by withdrawing the corresponding check-to-variable message from the a posteriori information
of the variable node. We note that this module concerns only messages sent to check nodes that are processed
by the next module and that these messages are updated only if they come from variable nodes that are not
leaf nodes with respect to the current iteration.
\end{list}
\end{mylist}

\begin{figure}[!t]
 \centering
 \includegraphics[scale=0.6]{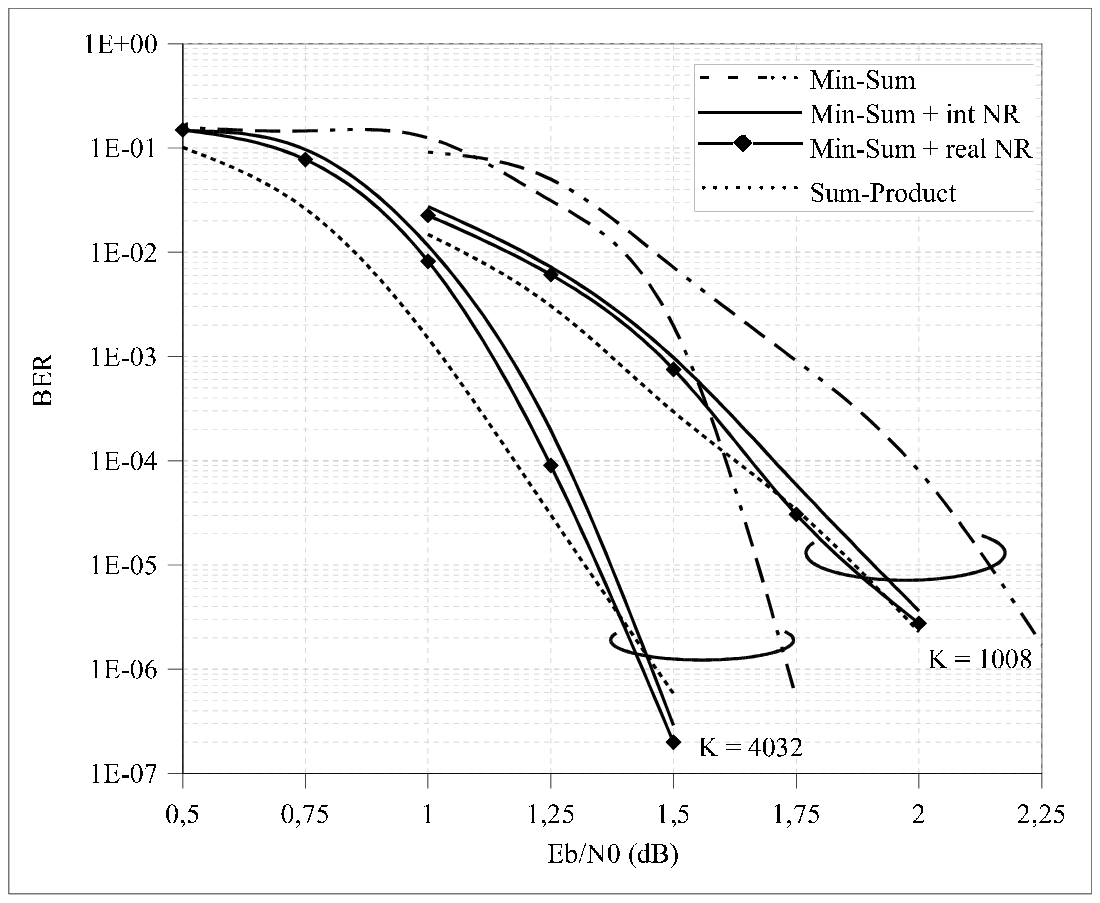}
 \caption{Performance of Sum-Product, Min-Sum and Min-Sum with Neighborhood Reliabilities decodings with binary LDPC codes (AWGN, QPSK,
 rate $1/2$, $K$ is the number of information bits, $200$ decoding iterations)}\vspace{-2mm}
 \label{nr_rate05}
\end{figure}

Figure \ref{nr_rate05} presents simulation results for irregular, rate $1/2$, binary LDPC codes. We observe
that for high signal-to-noise ratios the Min-Sum decoding with neighborhood reliabilities reaches the
performance of the Sum-Product decoding.

\begin{figure}[!t]
 \centering
 \includegraphics[scale=0.6]{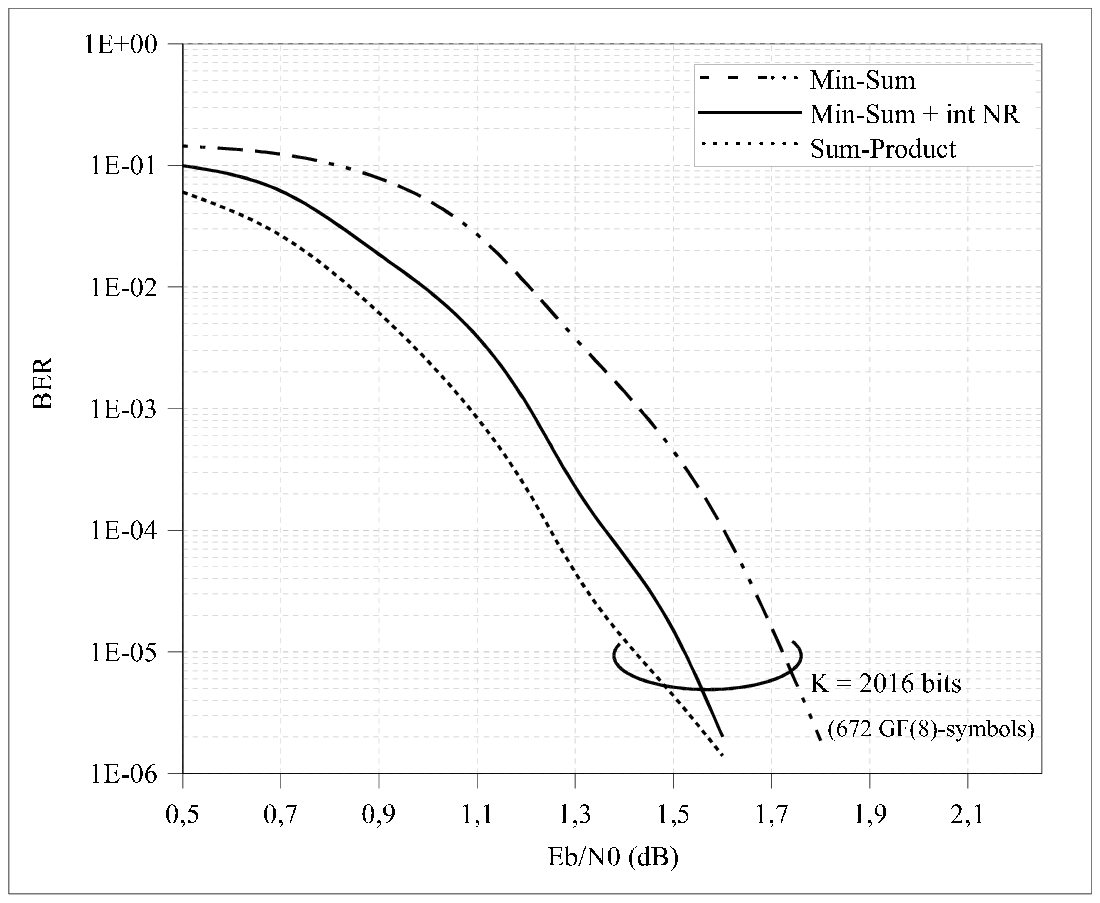}
 \caption{Performance of Sum-Product, Min-Sum and Min-Sum with Neighborhood Reliabilities decodings with LDPC codes over GF$(8)$ (AWGN, QPSK,
 rate $1/2$, $K$ is the number of information bits, $200$ decoding iterations)}\vspace{-6mm}
 \label{non_binary_nr_rate05}
\end{figure}

Neighborhood reliabilities can easily be generalized to the case of non-binary LDPC codes and the
integer-valued reliability function $\bar f$ can be used without any modification. Figure
\ref{non_binary_nr_rate05} presents simulation results for irregular, rate $1/2$, LDPC codes over $\mbox{\rm
GF}(8)$. Using neighborhood reliabilities the Min-Sum decoding performs very close to the Sum-Product
decoding at high signal-to-noise ratios.

\begin{rema}
We note that theorems $4.1$ and $4.2$ and corollary $3.1$ of \cite{Wiberg} also apply for the Min-Sum with
neighborhood reliabilities decoding. By using neighborhood reliabilities we only modify computation trees of
the a posteriori information.
\end{rema}

\begin{rema}
In case of non-binary codes, minimal deviations of computation trees can be
exploited in a completely different manner, leading to a self-contained decoding algorithm. This will be presented in a forthcoming paper.
\end{rema}

\section{Conclusions}
We showed that the processing order of nodes of a bipartite graph affects minimal deviations of different
computation trees and therefore impacts the Min-Sum decoding performance. This motivated the introduction of
neighborhood reliabilities and we developed a method that force minimal deviations to pass through reliable
nodes of computation trees. We proposed an integer-valued reliability function that can be easily integrated
to the Min-Sum decoding while preserving low complexity and independence on noise variance estimation errors.
At our best knowledge this is the first exemple of {\em self-correction} method for the Min-Sum decoding,
which applies to both binary and non-binary LDPC codes.

\bibliographystyle{abbrv}
\vspace{-1mm}
\bibliography{./bib/MyBiblio}

\end{document}